# Towards Automatic & Personalised Mobile Health Interventions: An Interactive Machine Learning Perspective

**Ahmed Fadhil**
University of Trento
Trento, Italy
fadhil@fbk.eu

**Yunlong Wang**
HCI Group, University of Konstanz
Konstanz, Germany
yunlong.wang@uni.kn

## ABSTRACT
Machine learning (ML) is the fastest growing field in computer science and healthcare, providing future benefits in improved medical diagnoses, disease analyses and prevention. In this paper, we introduce an application of interactive machine learning (iML) in a telemedicine system, to enable automatic and personalised interventions for lifestyle promotion. We first present the high level architecture of the system and the components forming the overall architecture. We then illustrate the interactive machine learning process design. Prediction models are expected to be trained through the participants' profiles, activity performance, and feedback from the caregiver. Finally, we show some preliminary results during the system implementation and discuss future directions. We envisage the proposed system to be digitally implemented, and behaviourally designed to promote healthy lifestyle and activities, and hence prevent users from the risk of chronic diseases.

## ACM Classification Keywords
H.1.2 User/Machine Systems: Human information processing

## INTRODUCTION
According to the World Health Organisation (WHO) poor diet and physical inactivity are public health issues and their associated health problems are getting beyond the healthcare capabilities [25]. These issues are strong contributors to overweight and obesity epidemic, which escalates to chronic diseases in the long-term. Research shows that the risk of developing chronic conditions can be reduced by adhering to a healthy lifestyle (e.g., a balanced diet and sufficient physical activity). There is a shift towards scalable solutions to promote healthier lifestyles outside the clinical settings. User adherence to the assigned plan is an indicator of the effectiveness of healthy lifestyle program. Yet, promoting deliberate lifestyle is not straightforward and maintaining a change in behaviour is a hard task to achieve. On the other hand, relatively few diet and physical activity applications have been tested in research environment to determine their effectiveness in health promotion [7]. Moreover, a small segment of such programs consider the delivery of practical and empathic health behaviour change support by considering cognitive, emotional and behavioural aspects of behaviour change. There is a need for a tailored system feedback that goes in parallel and fulfils users' preferences, while keeping the interaction simple.

Healthy lifestyle promotes optimal health and prevents health problems such as obesity and eating disorders [25]. This could also prevent long-term health problems, such as heart disease, cancer and stroke. There is a need to reinforce the adaption of long-term healthy eating behaviour. With this research we investigate formulating health promotion techniques with prioritisation based on user data.

To create sustainable healthy lifestyle, personalised feedback is always favoured over a one-size-fits all approach. Using modern sensor technology and proper algorithms, we can detect if a user is active, neutral or passive, and show dimensions of data about their activities. However, to guide users along their journey and create awareness, commitment to lifestyle goals it is necessary to offer interactive coaching support. Thus, based on the user activity data acquired, the caregiver entails the delivery of practical and empathic health behaviour change support, which is more personal and responds better to user feelings. A human in the loop can be effective in all lifestyle promotion domains, including physical activity exercise and food intake [31].

This paper provides an overview of an interactive machine learning to classify lifestyle promotion data. We first consider the state of the art view of trends in the field. For example, systems with goals and actions intended for health and wellbeing and applies some form of machine learning techniques. We highlight the importance and challenges associated with this emerging trend in lifestyle promotion. Finally, we discuss the case of CoachMe [11] a bot and web application for lifestyle promotion to a subject by comprising data corresponding to objective behaviour. We discuss the integration of an interactive machine learning algorithm into the system to classify users based on their activity performance.

## MACHINE LEARNING IN LIFESTYLE PROMOTION
With the increasing burden of sedentary lifestyle and overweight on our health, promoting healthier lifestyle becomes a necessity to prevent people from escalating into chronic diseases, such as obesity and diabetes. Developing systems to promote health and provide valuable information about user's habit becomes increasingly effective.

Machine learning is a fast-growing trend in the healthcare domain since it has the potential to be a powerful tool for human empowerment, touching everything from how we eat to how we diagnose diseases. Moreover, this can help health experts to identify trends that can lead to improved diagnoses and treatment, such as patient's health history and behavioural information data. Therefore, machine learning can identify aspects about user activities, such as behavioural pattern or the efficacy of the application. Understanding user preferences and observing their behaviour, we can interact with users at the right time, through the right channel, with the right tone, and the most relevant content. Machine learning can assist in developing more effective diagnoses and treatment, preventing prescription errors. This paper focuses on using a supervised machine learning algorithm to classify users and help caregiver to personalise their intervention feedback.

Combined human interaction into machine learning algorithms, the interactive machine learning highlights how to optimise human effort in machine learning model training [28]. Moreover, its methods are useful to analyse human behaviour and deduce health and wellbeing information. Recognising specific human behaviour analysis, such as eating pattern, daily physical activity are extremely import for healthcare providers to understand how to support their patients [3]. Moreover, human-machine collaboration is critical for the development of cost-effective and potentially cost-saving solutions. Companies like Google and Microsoft have partnered with a variety of healthcare organisations to implement machine learning solutions for complex problems, including medication adherence and cancer treatment [23]. These studies generally focus on human behaviour monitoring for health assessment purposes.

**BACKGROUND**

Interactive machine learning often involves complex iterations, where data is provided by an expert and then identify features to represent the data. Systems that learn interactively from end-users are becoming widespread. Involving users in such systems can increase the learning and ensure accurate output. Amershi et al., [4] presented case studies to demonstrate how interactivity results in a tightness between the system and the user. The paper also explores new ways for learning systems to interact with their users. Interactive machine learning is increasingly applied in social network to create custom groups. In another work by Amershi et al., [5] the authors discuss a novel end-user iML system "ReGroup" to help create on-demand groups in online social networks. The system interactively learns a probabilistic model and uses it to suggest additional members and group characteristics for filtering. Another work by Fogarty et al., [13] presented CueFlik, a Web image search application that allows end-users to quickly create their own rules for re-ranking images based on their visual characteristics. The study represents a promising approach to Web image search and an important study in end-user interactive machine learning.

Interactive machine learning provides huge support in health informatics together with the human-in-the-loop approach to solve computationally hard problems. Human experts can reduce an exponential search space through heuristics. In a study by Holzinger et al., [20] the authors discuss and evaluate iML and human-in-the-loop approach in enabling a human to manipulate and interact with an algorithm. The study selected the Ant Colony Optimization (ACO) framework and highlighted its importance in solving practical problems in health informatics, such as protein folding [21].

Interactive machine learning is an iterative process of running learner, analysing results, modifying data and repeating [18]. It has become the key components in several health and wellness applications. Several systems have been introduced to ease the process of building and deploying such applications. Microsoft released Azure machine learning, where the process is formalised as a data flow to ease applying machine learning algorithms to real world tasks [24]. Many types of interactive machine learning systems exist, we discuss systems applied in lifestyle promotion (e.g., healthier food choices and physical exercise). In this regard, Ge et al., [15] proposed a novel food recommender system to provide personalised recipe suggestions. The generated recommendations are in compliance with user preferences expressed by user ratings and tags which detects user preferred food ingredients. The study concluded that using tags in food recommendation can enhance prediction accuracy. For example, match the predicted preferences with the user's preferred recipes. Another work by Ge et al., [16] on food and lifestyle change developed a mobile platform to support people with healthier food decision making and reduce the risk of chronic diseases. The author stated that the application could be directly used in the kitchen and support meal decision making. An important step in healthy food recommender systems is the interaction design process which defines the user and system interaction. Therefore, according to Elahi et al., [9] a proper interaction design may improve user experience and result in higher usability. The work focused on the interaction design of a food recommender mobile app. The app captures user's long and short term preferences for food recipes. The long-term preferences are captured by asking user to tag familiar recipes, while user selects the ingredients to include in the recipe is used for short-term preferences. The app provides personalised recommendations based on these preferences.

A review by Clifton et al., [6] have introduced ways in which machine learning and software engineering could contribute to health informatics. The review discussed the contribution of both areas in health informatics domain. It is also important to highlight the correlation between machine learning and quantified self movement. Quantified self is a signal of the potential growth in the area of personal health data tracking. This movement includes user activists, such as self-tracked data on everything from diet and physical activity to results of medical test [22]. Based on the user tracked data, the algorithm can generate suggestion module in the form of feedback or indications about user performance. Ainsworth et al., [1] have developed MYBEHAVIOR that learns user behaviours via its machine learning model, then provides suggestions involve small changes to the existing behaviour. The study mainly focused on persuasion and behaviour change theories and provided users with common actions related to user's lifestyle and has been frequently done before [1]. A new version of MYBEHAVIOR application [29] used mobile sensing technology to develop personalised health feedback by combining behaviour tracking with recommender system algorithms. The system automatically learns user's physical activity and dietary behaviour and suggests changes towards a healthier lifestyle. The system used a sequential decision making algorithm, multi-armed Bandit [17], to generate suggestions that maximise calorie loss and are easy to adopt [29]. The result showed significant increase in physical activity and decrease in food calories.

Developing successful machine learning applications requires a substantial amount of data and algorithmic configurations. A study by Domingos et al., [8] summarised twelve key lessons learned when developing machine learning systems, which includes pitfalls to avoid, important issues to focus on, and answer to common questions. To develop an effective lifestyle promoting systems, machine learning could detect user emotional state and provide an appropriate feedback. However, users are heterogeneous, therefore it is extremely hard to target all user's emotional state with a single system. We believe a human expert in the loop could enhance system intelligence and effectiveness. The enhancement could be feedback on user activities and personalised recommendations. In the next section we discuss the possibility of integrating an interactive machine learning algorithm with CoachMe application [11] to promote user diet with a simplistic approach and a human in the loop. This model could be extended to other areas of personal health and wellness. For instance, preventive medicine, medication adherence and smoking cessation.

Applying a machine learning algorithm should be in compliance with behaviour theories to provide more effective recommendations and results. Behaviour analysis with learning theories assesses whether a person has the required skills to perform the behaviour. The next step is to increase or decrease the targeted behaviour. To illustrate, if a user is asked to go for a trekking trip, but the weather is cold or the user haven't done trekking before, the probability that he/she will follow the suggestion is very low [29]. With the ageing population and associated healthcare cost, machine learning can provide effective health or medical related recommendations to optimise healthcare cost and reduce the affluence of chronic people to care centres. Paez et al., [27] described a healthy lifestyle promotion system to provide valuable information about their habits. The system was developed around big data paradigm with bio-signal sensors and machine learning algorithms to provide a more personalised recommendations.

**PERSISTING CHALLENGES**

The current healthcare system lacks functions to help people preventing chronic diseases. Caregivers are short resource for providing personalised health recommendations for patients. The inherent coupling of the human and machine in interactive machine learning underscores the need for collaboration across the fields of human-computer interaction and machine learning. To solve these problems, intelligent software system is in demand. Lifestyle promoting applications integrated with machine learning can act as outlets for smart, persuasive feedback to the user and have a positive impact on healthy behaviour. These applications should address the challenges, including long-term engagement, contextualisation and individualisation, to provide optimal support for heterogeneous users. Unfortunately, no existing application can fit every user's need, hence providing different level of support becomes essential. Future work should ensure long-term usage of systems in health promotion and disease prevention, since short-term promotion cannot produce significant change in chronic conditions. Contextualisation and individualisation can be beneficial as they can make the system less intrusive and more efficient. For instance, users should choose what form of activity they would prefer (e.g., blend into the environment of user with reference to physical activities and daily diet). Another challenge is interdisciplinary collaboration when designing the mobile health intervention system, as there is no mature frameworks to guide researchers from different domain to design and implement the system.

**RESEARCH GOAL**

This paper discusses iML application and provides an overview of CoachMe application for lifestyle promotion. The application is focused on diet and physical activity tracking. This work highlights the following research questions:

- How to support caregivers to better understand patient's level of preparedness and provide a more tailored plan ?

- How to detect and classify patients based on their activity performance ?

- How can a human-in-the-loop be beneficial in lifestyle promotion ?

The use case scenario includes the endusers, the patients (e.g., person improving healthy habits) and caregivers who is limited to providing data, answering domain-related questions, or giving feedback about the learned model. The iML is expected to provide automation and detect users performance with respect to an activity.

We are mainly concerned with promoting healthier lifestyle through dietary recommendation and adherences. In such situation the patient suffers from poor dietary habits or barriers to adhere to a healthy lifestyle. To introduce changes in his lifestyle, it is advisable to focus on the preventive aspects of change, such as weight reduction, low diet in fat and sodium, and moderate physical activity [12]. However, we should begin by understating user preparedness to change. For example, their level of adherence and respond to various healthcare provided activities. Therefore, introducing machine learning techniques on top of such systems could significantly increase caregivers understanding about patients by visualising the outcome of their data.

This work contributes by developing a prototype system to promote active lifestyle and provide users with a simple tool (a Telegram bot application) to them and increase their self-motivation. The developed system with iML is based on the knowledge of user's preferences over a set of items. The input will be based on features that makes each user unique, namely their age, gender, BMI, and activity type they follow.

## COACHME PLATFORM: A CASE STUDY

### Background

This paper extends our previous work on CoachMe [11] which evaluated the state of the art from a behaviour recognition perspective and ways to promote long-term patient adherence to a healthier lifestyle. We intend to use a simple classification method to identify patterns in patient activity and classify patients based on similarity in their activity performance. This is to help the caregiver to discover how each patient performs with respect to a given activity and hence provide a more tailored activity that, according to the caregiver, better fits with patients' performance.

### CoachMe Architecture

The high level architecture contains the components that form the overall CoachMe system. The components and their interaction is represented in Figure-1.

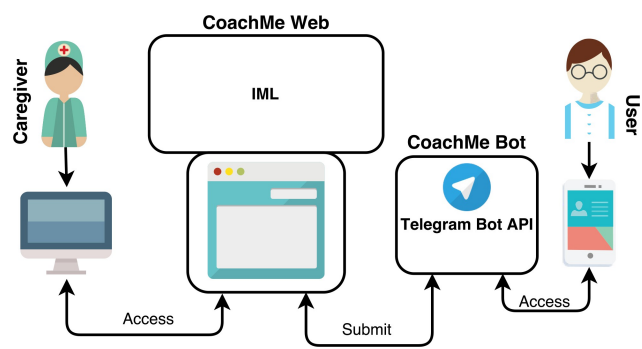

**Figure 1. The High Level Architecture Components of CoachMe.**

The architecture collects and processes data to promote diet and physical activity. With this in mind, we propose a simple and coherent activity monitoring solution that takes into account simplicity and non-invasive interaction. The architecture allows monitoring of non-chronic and healthy people, and is intended merely for lifestyle promotion and has no medical application within the study scope. From the technological point of view, the architecture consists of the following main components:

### Bot Application for Users

The bot application is an AI based conversational dialog engine which generates responses based on a collection of known conversations. The CoachMe bot application is a retrieval-based model which provides predefined. In this model, all the possible responses of the bot are predefined and rule-based. The bot uses the message and context of conversation to select the respond from a predefined list of responses.

The chatbot in this project provides user with a custom keyboard to access/report their daily activity using the Telegram Bot API. The user can view the dietary plan and submit a compliance about a certain plan with a single button click. The daily plans are predefined by the caregiver and there is a certain plan per user.

The output of this research will validate whether using bot for simplicity in user-caregiver interaction instead of auto generated recommendations will create a difference in adhering users to follow a healthy and sustainable lifestyle. In addition, by understanding participant's engagement with the bot, we could decide the targeted user group as potential candidates for evangelising the human-bot interaction. An example of typical daily plan on the bot UI can be seen in Figure-2.

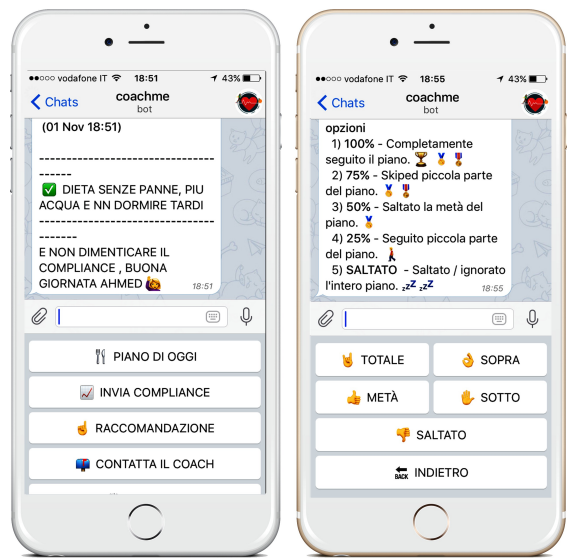

**Figure 2. A Dietary Plan Provided by CoachMe Bot.**

Each time a user clicks the new plan he will get a list of activities to follow. Afterwards, the user has to submit his compliance to each of the activities by clicking the compliance button. As the system receives more user compliance the user classification pattern increases. The data utility module can be used to train the chatbot.

### Web Application for Caregiver

The caregiver provides specific tailored activities to meet user's ability. The system provides timely notification through a Telegram bot application to trigger the user. The user can decide the amount of activity to follow, or skip the plan. The application will act as interoperable and is applicable to other domains in the context of lifestyle promotion and disease prevention.

### Messaging Platform

The web application allows the caregiver to deliver information to all involved patients in the system through the interoperability and messaging platform using push notification technology. On the other hand, the user can access the notification sent by caregiver via the Telegram bot application.

### Interactive Machine Learning Design

As shown in Figure-3, we create two prediction models for classifying user type. When a user first comes to the caregiver, he needs to register in the system by providing information, such as gender, age, BMI, education, and health condition. Then the caregiver provides a tailored plan for the user. We use the tailored plan as the label, as well as the user's profile as parameters, to train a KNN classifier, which should be updated once a new user's data is available. This is the process of creating the pre-prediction model. After the user starts to perform the provided plan, he needs to report his performance through the Telegram bot. The performance is stored and utilised as one of the parameters in the post-prediction model. When the caregiver decides to refine the user's plan, we get a label for this user. Thus we are able to train the second classifier.

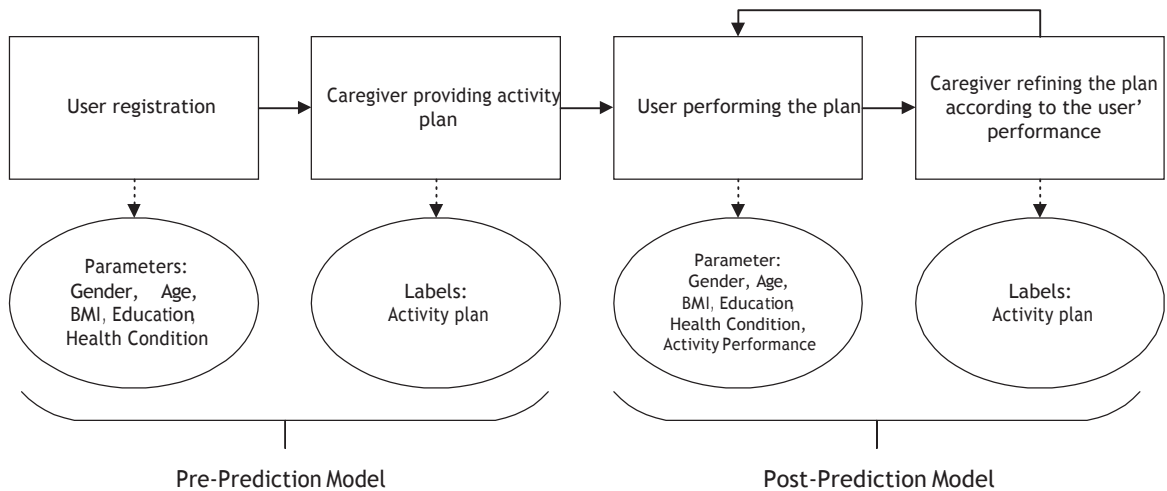

**Figure 3. The Interactive Machine Learning Design.**

### Caregiver Recommendation and User Feedback

The goal of caregiver recommendations and user feedback is to try and see the value provided by the solution, continues engagement, change behaviour, and maintain the behaviour change. Since even the simple act of tracking has been shown to have an impact, users can start with this simple behaviour and build upon it. The caregiver recommendation is the daily plan provided to the user, who can check it and later can provide his feedback represented by his activity performance.

### Caregiver

The caregiver has a significant role in adhering users to a healthy lifestyle. First, the caregiver provides tailored activities to meet user's ability. The system provides timely notification to trigger the user through a Telegram bot application. The caregiver can also interact with the user through a messaging window where they can communicate and exchange information about patient's condition. The system, using the machine learning model, will provide the caregiver with a ranking of patients on the system and tell the caregiver how their patients are performing with respect to the given activity.

The caregiver provides a weekly plan for the user. This plan consist of food and other activity suggestions. Moreover, the majority of these activities could be of the user's most frequent activities and the rest could be infrequent activities. The common user behaviour is used to promote user lifestyle in short-term. To target long-term health, the system explores infrequent activities that the user has to repeat in the future, leading to sustained activity.

### Activity Recognition

The system provides users with recommendations set by the caregiver who selects and compiles them from a predefined activity pool. The system provides personalised and contextualised activities based on specific user parameters and previous performance.

Activity recognition is an important component of CoachMe platform. Activities, such as walking more, eating more vegetables, or drinking more water can be accommodated. These activities are used to decide if the user is actively following an activity, is neutral and doesn't change, or is deteriorating and decreasing in terms of performance. The activity is labelled with the caregiver and contains certain degree of importance, later the activity is detected based on the given label.

### Temporal Detection

What is the best moment to trigger the user to perform a specific dietary activity. For example, the best moment to notify user to perform healthy dietary activity could be before meal preparation. Based on the activity type, the caregiver decides the application context and the best moment to trigger the user to perform it.

### Emotion Recognition

Human emotions have diverse effects on the immune system of the person and can have direct impact on the quality of life. To illustrate, positive emotions contribute to helping fight against cardiovascular incidents. On the other hand, negative emotions, for example, a high level of depression, may increase the risk of suffering from a stroke. That said, we plan to add a functionality for users to tell the caregiver how they feel at that moment. We will classify emotions as happy, sad, angry or neutral. This could support the caregiver with the recommendation and future activity assignments. Understanding user emotion is essential to provide them the right support, therefore, emotion recognition is a parameters within our iML model.

**Feedback Analysis**
The feedback mechanism can analyse users respond over the activity. The user provides the caregiver with their compliance data, based on which the system triggers a feedback and provides it to the caregiver. The feedback consists of user performance with respect to the activity.

**Suggestion Generation**
After classifying user behaviours, the system generates suggestions based on users past activities, this includes their food intake and compliance to the overall plan. The generated suggestion considers user's skills to perform a behaviour. For example, if the suggestion asks a user to go for bike ride, and the user has no access to a bicycle, the user will not follow the suggestion. On the other hand, if the user has performed a behaviour before, the skills can be assumed present.

**Theoretical Foundation**
This study references the Fogg's Behavioural Model to apply theoretical principles into technology design. With CoachMe system, we focus on promoting low-effort actions that can be triggered even when motivation is low [14]. Thus, CoachMe suggests (e.g., cues or triggers) a frequent behaviour (e.g., a particular walk) that the person often does in a particular life context. This can increase the frequency of behaviour that a person already does. However, the system with the caregiver can suggest a new behaviour (e.g., go jogging) that would burn more calories and the person is able to perform. In order to successfully sustain such behaviour it has to be repeated frequently until it becomes a habit [14].

**User Types**
Recent approaches on user motivation and adherence towards healthy lifestyle have witnessed an increasing number of applications that treat users as a monopolistic group in their design. This is a bad strategy since an approach that works for an individual may actually demotivate behaviour in others. With this work, we develop a model to classify users as Active, Neutral, or Passive, based on their overall performance and other parameters. This classification is based on the gamer's type strategy discussed by Orji et al., [26]. We will employee personalisation approach to better persuade a particular type of users.

**Activity Clustering**
On the web application side, the caregiver creates and clusters similar activities together. For example, similar food items are clustered based on their ingredients. Based on user feedback, the system could detect if the user is repeatedly having high-calorie intake or skipping the given plan and cluster him together with other users with the same level of adherence. Based on user's daily diet, CoachMe system will provide the user with an activity that matches with their ability which could be an infrequent activity. The user will take up some of these infrequent dietary activities and make them frequent in the future.

**Model Evaluation**
This work adapts an interactive machine learning (iML) where there is a human involved in the loop represented by the caregiver. There is evidence that humans often still outperform machine learning techniques [19]. For example, a promising technique in diagnostic radiologic imaging to fill the semantic gap is to adopt an expert-in-the-loop approach, to integrate the physicians high-level expert knowledge into the retrieval process by acquiring his relevance judgments regarding a set of initial retrieval results [2]. One drawback of iML-approaches is that methodologically correct experiments are very difficult to replicate, since human agents are subject and hence cannot be copied in contrast to data and algorithms. However, still iML could help equip algorithms to support caregivers in understanding various user behaviours and adherences to the given activity. The importance of iML becomes apparent when the use of hybrid solutions is not enough or difficult.

**Scenario Description**
To better understand how the system works We provide a scenario describing user interaction with the system. Consider the healthy dietary adherence scenario for people, like John. He is a 40 years old man, who wants to adapt an active lifestyle in his daily routines. John would like to eat fruits and do physical activities, such as walking and hiking. Recently John have developed the habit of eating fast food for his dinner. This led to weight gain and has affected his physical activity and overall lifestyle. Furthermore, John also has the routine of socialising with friends and he is an active user of messaging platforms, such as WhatsApp[1] and Telegram[2].

*John is interested to know about his health primitives and improve his lifestyle. He is looking forward to change his poor dietary habit and physical inactivity. He is looking for a tool to tell him about his physical activity in the previous days and the dietary pattern he follows. Furthermore, John is interested in a simple and non-intrusive tool that provides him with real recommendations by checking his daily conditions. He is not just interested in some system generated feedback based on his performance. John would like to receive weekly plan and follow a routine daily plan to achieve his active lifestyle goal. Supporting this scenario means avoiding hybrid solutions at least for the phase regarding user weekly plan. Moreover, the system should allow the user to report his daily activity in a truly ubiquitous and simple manner. In addition, the user is not obliged to provide amount of calories burned and hours of workout. He can report all his activities by a simple few button clicks. John can access the CoachMe bot application directly*

---

[1] https://www.whatsapp.com/
[2] https://telegram.org/

*from his Telegram application. With just few clicks, he can register to the bot and start receiving updates. CoachMe provides also a web application interface that is mainly for caregivers to assign, track and monitor Johns condition. The web interface gives access to visualise the user's daily, monthly and weekly physical routines through our graphical widgets. Now the caregiver can provide John with a weekly plan that contains daily activities. John can follow the daily activities and at the end of the day he is triggered to submit his compliance. By submitting sets of activities the iML classifier will categories John among a certain category (either active, neutral or passive). In this way, when the caregiver accesses the web interface he can check and see John's performance with respect to each plan. The classifier considers features, such as users' age, gender, height and weight, along with the user's preferred physical activity and food.*

The recommendations mechanism by caregiver is in accordance with user's feedback about the provided recommendations. This work has a great potential for promoting an active lifestyle to improve individual as well as the population's health.

**PRELIMINARY RESULTS**
The machine learning model chosen in this study will be mainly for decision making. It is to classify users with unknown classes and based on sets of rules or types of models it classifies the new user into existing classes. This could support caregivers with valuable information about the habits and daily patterns of a user and permits the caregiver to recommend a more tailored activity to the user. This recommendation system can be used as enabler in health intervention bringing some new functionalities. For instance, based on users activity data, a recommender system can guide him about the necessary actions needed to be taken.

So far, we have developed the prototype system and currently working on integrating the iML part into the system. We are collaborating with caregivers at an ambulatory clinical centre and UX designer and the prototype design was build based on their feedback. For example, the caregiver suggested the inclusion of activity clustering into the system to form a weekly plan, whereas initially it was for a single activity and for a daily plan. Moreover, both the caregiver and UX designer perceived the inclusion of Telegram bot as positive. Since most patients are already active users of Telegram application or similar chatting applications, which removes the barrier of introducing this technology. Hence, we ensure the consistency of the work from medical and design point of view. The system considers various types of recommendations that ranges from:

- Food suggestions on the best dietary changes to improve user health.
- Recommendation for harmful substances, such as drugs and smoking.
- Recommendation for relaxation, such as yoga.

This work contributes a new approach to social access control using end-user iML to help caregiver track and provide personalised recommendations to their patients. The interaction cycles in iML are more rapid, focused, and incremental than in traditional machine learning. This increases the opportunities for users to impact the learner and, in turn, for the learner to impact the users. As a result, the contributions of the system and the user to the final outcome cannot be decoupled, necessitating an increased need to study the system together with its potential users [4]. Formative user studies can help identify user needs and desires, and inspire new ways in which users could interact with machine learning systems. User studies that evaluate interactive machine learning systems can reveal false assumptions about potential users and common patterns in their interaction with the system.

**DISCUSSION**
The CoachMe system has as goal the design and development of a platform targeting people at all stages of disease continuum (health promotion, managing prevention, managing at risk, and managing complications) to increase their quality of life before more acute episodes. In this work, we have focused on the main architecture description with details about the various components involved and some preliminary results to be considered to demonstrate the functionalities of the developed architecture. We discussed both the iML classifier and the inclusion of Telegram bot in the interaction. This is an effective way to adhere users to a long-term plan. Users spend the vast majority of their time in apps that provide them with chatting services. The reason is because using conversation is easier than heavy interfaces in the application [30]. It is crucial to create a user experience that engages and satisfies users need quickly.

More work is needed to improve the training and test datasets and find a better way to train the model. Finally, social network and gamification are also two important aspects that should be considered in future work, since they can increasingly contribute to the motivation and user engagement. Hence, future work will pay particular attention on integrating some aspects of these two techniques into the CoachMe platform.

**CONCLUSION**
Interactive ML is becoming increasingly involved in health informatics and lifestyle promotion, as there is more need to aid caregivers with user behaviour and performance. In this paper we presented machine learning application in lifestyle promotion, we then discussed the high level architecture of CoachMe system with details about the components and preliminary scenarios to demonstrate the functionalities of the architecture.

We discussed the application of iML machine learning algorithm to detect user activity performance and classify users accordingly. As a preliminary result, we developed a web application for the caregiver and provided users with a Telegram bot application to report their daily compliance. The iML model is a promising classification approach to detect active, neutral and passive users based on their performance. The classifier is part of the web application to identify users with high or low adherence to common recommendations given by the caregiver.

The major strength of this study is combining the iML algorithm with Telegram bot and a human actor in the loop to achieve a personalised intervention. Future work should focus on testing and validating the approach with real users. Finally, interdisciplinary work between medical professionals, engineers and psychologists is key for the success of such applications.